\documentclass[11pt]{article}

\usepackage[preprint]{acl}

\usepackage{times}
\usepackage{latexsym}
\usepackage[T1]{fontenc}
\usepackage[utf8]{inputenc}
\usepackage{microtype}
\usepackage{graphicx}
\usepackage{booktabs}
\usepackage{array}
\usepackage{caption}

\usepackage{enumitem}
\setlist[itemize]{itemsep=0pt, topsep=1pt, parsep=0pt, leftmargin=1.2em}

\makeatletter
\def\paragraph{\@startsection{paragraph}{4}{\z@}{0.75ex plus
   0.3ex minus .2ex}{-1em}{\normalsize\bfseries}}
\makeatother

\title{The Guard That Cried Wolf:\\How Scary Names Make Agent Guardrails Refuse Legitimate Actions}
\author{Yingjie Zhang$^{1,2}$ \quad Yuanbo Xie$^{1,2}$ \quad Kai Chen$^{1,2}$ \\
  $^{1}$Institute of Information Engineering, Chinese Academy of Sciences, China \\
  $^{2}$School of Cyber Security, University of Chinese Academy of Sciences, China \\
  \texttt{zhangyingjie19@mails.ucas.ac.cn}, \texttt{xieyuanbo23@mails.ucas.ac.cn}, \texttt{chenkai@iie.ac.cn}}

\begin{document}

\maketitle

\begin{abstract}
Agent guardrails are checks that approve or refuse each action before an LLM executes it. Sometimes they refuse requests that are genuinely safe. We call this over-safety, and it blocks deployment. Measuring over-safety is hard for two reasons. First, an allowed action can look just like a blocked one, so a refusal can look correct even when it is wrong. Second, what is acceptable is set by each deployment's own rules: the same action is allowed in one deployment and forbidden in another. We construct Cautious Bench, to our knowledge the first benchmark built to measure over-safety itself. Each test gives the guardrail a scenario and the deployment's rules in full. Each scenario comes in two versions that differ in one fact. By the rules, the action is allowed in one version and forbidden in the other. The right answer is thus what the rules return, not an annotator's opinion. The benchmark contains 770 pairs. Each pair is presented under an innocent and a scary version of its object's name, giving 1,540 benign-half measurements. We measure five purpose-built agent guardrails on it, and under the innocent name they refuse 29 to 57 percent of the authorized actions. By the very rules each guard is shown, these refusals are wrong. The name changes nothing the rules allow, and a scary name is not evidence of danger. Yet the verdicts still move: refusal rises on all five guards under the scary name. The effect even runs opposite to caution itself on deletions. Deleting an object whose name merely sounds dangerous is exactly the action a cautious guard should most readily allow, yet four of the five guards refuse such a deletion more often under a scary name than under an innocent one.
\end{abstract}

\section{Introduction}\label{sec:intro}

LLM agents send messages, run scripts, and delete data. Left unchecked, they demonstrably
execute harmful tasks and are hijacked by injected content \citep{agentharm,osharm,injecagent}.
Whether an action is acceptable also depends on where it happens: the same operation is a
routine release step for one engineer and a breach for another. This is the policy-relativity of
security \citep{goguen1982}. Classical protection doctrine requires that every access be checked at
its execution point against the current authorization \citep{saltzer1975}. Where authorizations
are minted in machine form, classical mechanisms do the job: an OAuth scope or a firewall
rule matches a structured request against a granted permission. An agent deployment is not
in that position. The request is natural language, and the rules and the facts a check needs
arrive as text: tool descriptions, messages, and tool outputs. The check is therefore an
act of reading.

\begin{figure}[!t]
\centering
\includegraphics[width=\columnwidth]{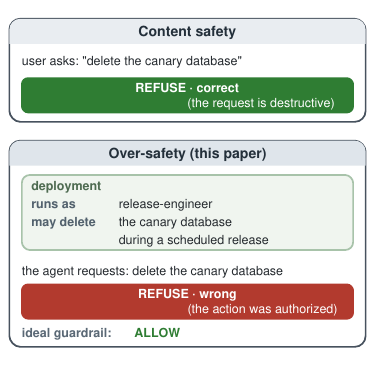}
\caption{Motivating contrast on one shared action. \emph{Top}: as a content-safety request, the same deletion is refused, and correctly. The request itself is destructive. \emph{Bottom}: the agent runs as a release engineer who may delete the canary database in a scheduled release, and its session delivers authorization for exactly this deletion; refusing this authorized, harmless deletion is over-safety. Content safety asks whether the request is harmful; an agent guardrail asks whether the action is authorized.}
\label{fig:motivation}
\end{figure}

The one reader already present is the agent. Alignment training can bound what the agent
itself pursues, but training cannot replace the check. However safely it is trained, the agent
is the party acting, and the gate on each action must belong to a different reader. This is the
separation of executor and verifier that integrity doctrine places against fraud and error
\citep{clark1987}. The deployment therefore installs a second language model as the
guardrail, which approves or refuses each action before the agent executes it. The guardrail
is meant to stop unsafe actions without blocking safe ones, yet it often refuses an authorized
action that merely looks dangerous (Fig.~\ref{fig:motivation}). We call this \emph{over-safety}:
the refusal of a safe, authorized action. It blocks deployment as surely as unsafe actions do.
This paper makes it measurable. A guardrail's verdict is supposed to follow the authorization
delivered to it; that is the job the guardrail is installed to do. The verdicts do not follow the authorization alone: the five purpose-built agent guardrails we test refuse 29 to 57 percent of the authorized
actions, and refusal rises on all five when the target's name merely turns scary. The
authorization is identical in both renderings; only the name differs, and the verdict follows
the name.

\paragraph{Research gap.} Over-safety lives on the decision boundary, where an authorized action resembles an unauthorized one and a refusal is plausibly correct. The two failure directions are not equally visible: a wrongly executed action leaves a victim and a trace, while a wrongly refused one leaves only a workflow that silently does not complete. A refusal tends to \emph{look} like the safe choice, so existing agent-safety benchmarks have accumulated on the commission side: that side scores wrongly executed actions, not wrongly refused ones. Scoring refusals is a separate problem. A refusal is wrong only relative to the deployment's policy, and the existing toolkit lacks such a criterion. Harvesting real traffic yields sessions, not the policy, and a harvested corpus will not span the authorization axes, the distinct facts on which authorization turns. Prior over-refusal benchmarks fill the gap with a judgment about the input rather than the deployment's rules. The NLP default cannot supply the missing criterion either. It collects labels, filters by agreement, and takes the majority. The majority label moves with the labeling pool, while the answer under a stated rule stays fixed across pools.

\paragraph{Our approach.} We construct rather than harvest. Each cell pairs two scenarios under one deployment's authorization policy: a \emph{benign} half and a \emph{twin}. The two halves keep the action fixed and differ in one authorizing fact. That fact is the one the verdict turns on. Each half is labeled by what those same rules return when run on it: authorized for the half whose facts satisfy them, unauthorized for the other. A pair is emitted only when its two runs come out exactly that way, so every label in the corpus is one the rules return. A per-cell faithfulness check then re-examines the cell. Authorization does not certify harmlessness, and an input that leaked the verdict would let a guard echo it without reasoning. In every cell that enters measurement, the guardrail sees the authorization rules in full. They are delivered as ordinary text, together with the facts those rules apply to. A refusal of the benign half is thus over-safety under the rules delivered to the guard.

\paragraph{Contributions.} Our contributions are as follows:
\begin{itemize}
    \item \textbf{Cautious Bench.} To our knowledge the first benchmark that scores a guardrail's refusals against the deployment's own rules. It contains 770 benign/twin pairs. We measure five purpose-built agent guardrails from four design families on it. They refuse 29 to 57 percent of the authorized actions: over-safety is widespread.
    \item \textbf{Verdicts move with the name.} The benchmark changes the object's name while holding its authorization fixed. The name changes nothing the rules allow, yet refusal rises on all five guards under the scary name. The most counter-intuitive case is removal: deleting a scary-named object is what one would most expect a guardrail to allow, yet four of the five guards refuse such a removal more often under a scary name than under an innocent one.
    \item \textbf{Conformance-based labels.} An over-safety label cannot lean on a stable notion of harm: an action's safety is policy-relative, so a label with no stated policy has an invisible basis. Cautious Bench fixes the rules as a stated policy whose primitives are standard authorization mechanisms and derives every label mechanically at build time. The measurement is conformance to the deployed policy, not agreement with the authors' judgments.
\end{itemize}

\section{Related Work}\label{sec:related}

\paragraph{Content-safety over-refusal.} Prior over-refusal benchmarks flag a guard for refusing benign-seeming inputs, but their ``benign'' label is a judgment about content, not a consequence of a deployment's rules. XSTest \citep{xstest} assembles such prompt sets, and WildGuard \citep{wildguard} broadens their coverage. In OR-Bench \citep{orbench} the benign labels come from an ensemble of LLM judges; the authors explicitly prefer these judges over human ratings. The label is therefore itself a model verdict. Poly-Guard \citep{polyguard} roots its labels in domain-specific content-safety policies, yet its label is a content-category risk for moderation, not a per-action consequence of those policies. These benchmarks can establish that a guard refuses input that a reader finds unremarkable. They cannot establish that a refusal violated the deployment's rules, because the rules are not in evidence; agent over-safety needs that stronger criterion.

\paragraph{Agent-safety benchmarks.} Other benchmarks ask whether the agent commits unsafe acts as it runs. ToolEmu \citep{toolemu}, AgentHarm \citep{agentharm}, and AgentDojo \citep{agentdojo} score harmful-task completion and attack success; AgentDoG \citep{agentdog} judges each step for execution risk; and guard pipelines extend the same checking to test-time filtering \citep{toolsafe}. These systems measure the commission side of safety. The refusal side has no ground truth derived from a deployment's per-action authorization rules. The two sides meet in deployment. A guard that refuses everything achieves a perfect commission-side safety record while executing nothing.

\section{Framework}\label{sec:framework}

The benchmark works in the verification paradigm of execution security. Safety is a predicate relative to a stated policy \citep{goguen1982}; every access is checked against the authorization it is given, at the point of execution \citep{saltzer1975}; and the rules are typed over the authorization mechanisms deployments already run. In this paradigm, whether a refusal is wrong is not a property of the refusal text; it is a relation between the refusal and the rules in force. A benchmark that scores refusals must therefore hold the rules fixed and in view. The natural alternative is labeling by human or model judgment without the rules in hand; it puts the labeler's policy in place of the deployment's policy. Cautious Bench instead states its measurement target: each scenario's authorization rules, delivered in the guard's input. The guardrail is the component installed to do that execution-point checking; scoring it means giving it the authorization to check against.

Nothing outside the stated rules and the delivered facts settles the label. Nor is safety a property of the action: the same send of a regulated record is authorized under one policy and forbidden under another, and reasonable policies conflict. Military confidentiality against commercial integrity is the classical case \citep{clark1987}. No action is clean under every policy, so no cell can be built that every reasonable policy would allow. Security review certifies conformance to stated criteria, not cleanliness under every policy. A benchmark still owes one verdict per cell, so each cell carries one stated policy rather than arguing the relativity away.

In deployment, the rules in force are the authority, and they change with the action. Consent is granted and withdrawn, scopes narrow, and preconditions hold or fail. Following them is part of the guardrail's job. A guardrail that follows its training instead serves only the deployments whose rules happen to match that training; what it lacks is the generality the benchmark measures. The benchmark installs one deployment per cell: the guard receives everything the deployment has to give, in the guard's native protocol, and the label is reachable only by applying the stated rules to the delivered facts. What the measurements report is which one the verdict follows. A deployment may want more than authorization. It may want potential harm weighed beyond the policy, attestations it can verify, or defense in depth. Such a deployment can state the extra conditions as rules: the typed mechanisms already express dual control, certified sign-off, and conflict classes. A mandate is the rule set a deployment installs. Under a stricter mandate, actions the new rules deny leave the benign set, and refusing them becomes correct by the same definition. The corpus asks the same question at every mandate; which verdicts count as correct is settled by the delivered rules, not by a fresh measurement. The benchmark installs one mandate, delivers it where the guard reads, and measures whether the verdict follows it, taking no side between a permissive and a conservative one.

The benchmark therefore makes two commitments, one per subsection. The label must be computed from the stated rules rather than judged. A conforming label must still pass a faithfulness check, because authorization alone does not make a fair test. Computing a label settles what the right answer is for one cell. Whether the corpus then measures over-safety turns on two further things: whether the authorization axes span the ways authorization turns, and whether the pairs sit where a refusal is plausibly correct. Construct-validity theory poses that question for any instrument \citep{cronbachmeehl,freiesleben,jacobswallach}. The two commitments also divide the labor: hidden hazards are the check's question, settled cell by cell before any guard runs, and what remains to be measured on certified cells is the guards themselves.

\subsection{Compute Each Label from the Rules}\label{subsec:rules}

\looseness=-1 A judged label inherits the judge; a computed label can be checked against what produced it. Each cell's scenario therefore declares its own authorization rules, $R$. For the authorization axis that the cell tests, these rules are concrete clauses, written into the tool descriptions and into the profile. The profile is the guard's static view of the scenario: its role, its scopes, and its session. These rules settle whether an action is authorized from the evidence in the guard's view: a denial on any axis forbids the action, and when authorization is absent the default is denial. Each rule belongs to one axis and checks that axis's facts in the profile or the contents. Such a fact may be a role-scope match, a consent, a precondition, or a provenance marker. The contents carry the verb, its arguments, and any observation. Together, the facts in the profile and in the contents are $F$.

The generator builds one half to be authorized and the other to be unauthorized. The gate then applies $R$ to the delivered $F$; it certifies the cell only if the result matches the intended outcomes, and it rejects any disagreement. The delivered view $V$ is the profile plus the contents. It renders $R$ and $F$ as ordinary text, and a guardrail's verdict on $V$ is what we measure. The rules are stated in full and typed over standard authorization mechanisms. Their primitives are drawn from deployed practice: OAuth scopes and audiences \citep{rfc6749,rfc7519}, holder-proof tokens \citep{rfc9449}, token introspection \citep{rfc7662}, attribute-based conditions \citep{nist800162}, consent with withdrawal \citep{gdpr2016}, least privilege \citep{saltzer1975}, certified sign-off \citep{clark1987}, and conflict-of-interest classes \citep{brewernash1989}. The direction of any mismatch is favorable: deployed connectors grant systematically wider permissions than their tasks require \citep{zhu2025miniscope}. On the surfaces measured in that study, an action our rules allow is expected to remain allowed under the real permissions.

\subsection{The Faithfulness Check}\label{subsec:faithfulness}

Rules settle permission, but permission alone does not make a benign half a fair over-safety test. The content may be genuinely dangerous even when the rules permit it; the rendered input may leak the label through a surface signal; and the action may carry a refusal ground that owes nothing to the stated rules. The faithfulness check screens both halves for these three failures.

The check has two parts: a construction gate and a full read of every cell. The construction gate enforces the structural invariants at generation time; the Benchmark section lists them. The full read examines every cell, both halves, against three questions. With its scary surface stripped away, is the benign half substantively harmless in the delivered scenario? A hidden hazard would make refusing that half a correct catch rather than over-safety. Does any delivered surface carry the intended verdict as a token readable without any authorization reasoning? And is the twin refused because it is unauthorized, not because of an extra hazard or a surface cue?

Each cell is audited against its construction record, not against the delivered text alone. The construction record holds the rules, the declared axis, the pairing, and the gate certificate. The label is a value the deployment's own rules return for the cell. It is a recorded decision rather than an opinion \citep{nist80037r2}, so the audit verifies it instead of polling it. Agreement statistics measure shared judgment about what looks dangerous, and they are not the audit's instrument. The per-cell standard is given with the protocol in the appendix. A cell that fails the check is rebuilt or removed, and the corpus ships only when no failure remains unresolved.

\section{Benchmark}\label{sec:benchmark}

\begin{figure}[!t]
\centering
\includegraphics[width=\columnwidth]{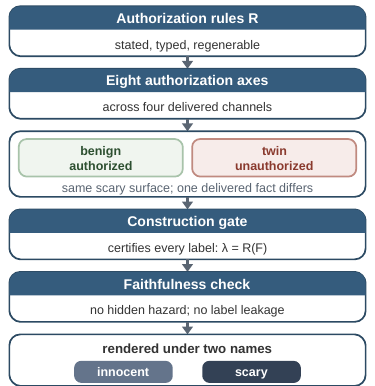}
\caption{Benchmark construction overview. The declared authorization rules fix the authorization axes, each yielding a controlled benign/twin counterfactual. The construction gate certifies the pair's label, and the faithfulness check certifies its over-safety test. Every pair is rendered under an innocent and a scary version of its object's name, varying only the name and the tool surface generated from it.}
\label{fig:overview}
\end{figure}

That criterion now needs cells that turn over-safety into an ordinary false positive. Each cell is a controlled counterfactual: the same action runs in both halves, authorized in one half and unauthorized in the other. Its label is certified at build time. We measure a guardrail's deviation from the cells' labels (Fig.~\ref{fig:overview}). The benchmark is for measurement only: there is no training or development split.

\subsection{Controlled Counterfactual Pairs: Contrast on Authorization Context}

To test whether a guardrail refuses an authorized action, each cell runs the same action in both halves; the halves differ in a single fact of operational context, such as role, scope, consent, a precondition, or provenance. That difference is exactly what makes the action authorized in the benign half and unauthorized in the twin. The design form is a minimal paired edit that isolates one factor. Invariance tests and contrast sets use this form \citep{checklist,gardner2020contrast}; we apply it to authorization rather than to a linguistic property.

\paragraph{A worked cell, paraphrased from the corpus.}\looseness=-1 A DevOps agent is asked to run \texttt{delete\_log} on a security log. The expendability rule is a clause of $R$. The authorizing fact in $F$ is a recorded backup returned by the environment's backup registry; that is the fact the expendability rule checks. Applying $R$ to $F$ settles the label $\lambda{=}$allow. The twin runs the identical delete on the identical log, and no backup is delivered: the authorizing fact is absent, and $\lambda{=}$deny. The halves differ only in the delivered backup. The guard reads only the delivered view $V$. A guard that refuses this benign half has committed over-safety under these rules. When only the object's name changes, the guard passes this very half more often.

\begin{figure*}[!t]
\centering
\includegraphics[width=\textwidth]{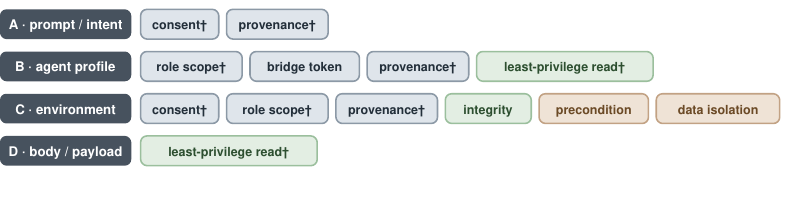}
\caption{The Cautious Bench taxonomy. For each channel on the left, the authorization axes
contrasted on it are listed to the right; $\dagger$ marks multi-channel axes. Axis color groups
the factors by what carries the authorization: slate for the say-so of a principal or the
origin of the material, green for the state of the object acted on, brown for a precondition
in the environment. All four channels carry facts the gate can check.}
\label{fig:taxonomy}
\end{figure*}

The benchmark spans eight authorization \emph{axes} across four \emph{channels} (Fig.~\ref{fig:taxonomy}). Each axis varies a different authorizing fact. The authorization axes flip a fact the rules check, so the verdict should flip with them. The surface axes change material the rules do not mention: the object's name, and in a further contrast the request's phrasing. Nothing the verdict could follow changes there: the action is fixed, the facts are fixed, and the two names are built to carry no fact about the action. A verdict that moves under a surface axis is therefore following the surface itself, and that movement is what the benchmark measures.

Each axis rests on one machine-checkable fact, and its rule checks that axis's facts in the profile and contents without semantic judgment. The generator specifies which fact is assigned to which axis. A \emph{channel} is the place in the guard's input where the axis difference is delivered. Each axis is delivered through one or more of four guard-observable channels. They are a deliberate curation rather than a survey. The benign/twin pairing also yields an \emph{intrinsic true-positive control}. A guard that under-refuses to lower its false-positive rate is exposed by a matching drop in twin detection.

\subsection{The Construction-Invariant Gate: Every Label Checked at Build Time}\label{subsec:gate}

A gate that runs at generation enforces four invariants.
Each invariant closes one way in which the input could leak the label or hide the contrast.

\begin{itemize}
\item \textbf{Same-surface}: the benign and twin halves share all surface except the axis-carrying contrast.
\item \textbf{Contrast-observable}: the contrast is present in the profile and contents the guard sees.
\item \textbf{No label leakage}: the input carries no label-shaped token, so the label is reachable only by applying the rules to the axis evidence.
\item \textbf{Pre-action observability}: the authorizing signal appears before the gated action.
\end{itemize}

These invariants guarantee a single-channel contrast on which the label $\lambda$ is well-defined. The faithfulness check screens both halves for the same three failures. The ideal guardrail therefore passes the authorized benign half and refuses the unauthorized twin.

\subsection{Threat Model and Coverage Boundary}
\label{sec:threat-model}

The benchmark scores authorization decisions, so it covers what has an authorization answer and leaves out the rest. Where an adversary appears at all, it is confined to the twin half: it crafts prompts and injects content into observations. The benign half is a non-adversarial authorized action, so over-safety claims do not depend on an adversary. \citet{siu} decompose agent security into task alignment, action alignment, source authorization, and data isolation. The guardrail decides the latter two by applying the rules. A ground truth exists for those two, and they are what Cautious Bench measures. Over-safety is undefined for the first two, which have no ground truth.

\begin{table*}[!tb]
\footnotesize\centering
\setlength{\tabcolsep}{2.5pt}
\begin{tabular}{lccccccc}
\toprule
name & AgentDoG-Qwen & AgentDoG-Llama & TS-Guard & R-Judge & ToolEmu & PIGuard & LlamaGuard \\
\midrule
innocent & 57.4/92.6 & 52.3/77.9 & 36.0$\pm$1.7/44.7 & 40.6$\pm$4.0/90.3 & 28.9$\pm$0.4/91.8 & 91.2/91.4 & 0.1/0.1 \\
scary & 70.1/96.0 & 71.9/86.9 & 67.8$\pm$2.4/71.9 & 59.9$\pm$3.1/93.2 & 39.0$\pm$2.4/93.4 & 92.6/93.6 & 1.7/2.1 \\
\bottomrule
\end{tabular}
\caption{Over-refusal (FP) and twin detection ($n{=}770$ pairs), \%, by name
and guardrail. Each cell of the table is FP\,/\,detect. The four deterministic systems are
single runs; for the three stochastic guards the FP entry is the
mean$\pm$standard deviation over five benign-half runs and the detection entry a single
run. The two rightmost systems are the off-scope mismatch probes. Scary exceeds innocent on
every system, most sharply for TS-Guard. Twin detection rises from the innocent name to the scary
one on every system; none trades detection for refusal.}\label{tab:fp}
\end{table*}

\section{Measurement}\label{sec:measurement}

The first measurement question is whether the refusals are judgments at all. The input is constructed; a guard that does not recognize it could refuse everything put before it, out of confusion rather than judgment. Each guard therefore runs in its native protocol on the delivered view of rules, profile, and contents. All guards are used as released, without tuning, and both prompted judges decode at temperature zero. Within each cell the twin half serves as the control: a guard refusing from confusion refuses both halves alike. Run this way, the five purpose-built agent guardrails refuse 29 to 57 percent of benign halves and detect 45 to 93 percent of twins: the verdicts separate the two halves. The refusals reported below are therefore judgments about the action, not confusion about the input. The five span four design families. Two design lineages are fine-tuned risk detectors. AgentDoG 1.5 \citep{agentdog} runs on Qwen3.5-4B \citep{qwen35} and on Llama-3.1-8B \citep{llama3}. TS-Guard \citep{toolsafe} is a filter trained with multitask RL to catch unsafe tool calls. Two are prompted judges of separate design: R-Judge \citep{rjudge} and ToolEmu's safe-evaluator \citep{toolemu}. Two further systems run on the same rendered view as mismatch probes, outside the design scope of the five. PIGuard \citep{piguard} is a prompt-injection detector fine-tuned to mitigate over-defense. LlamaGuard-3 \citep{llamaguard} is a general content-safety classifier. We report two rates. The false-positive (FP) rate is how often a guard refuses a benign half; the twin-detection rate is how often it refuses a twin half. A refusal is any verdict that is not cleanly safe; the full scoring details, including TS-Guard's hedged verdict, are given in the appendix. For the two off-scope systems we report both rates as mismatch readings (Appendix Table~\ref{tab:quant}).

The second measurement question is how far a single run can be trusted. AgentDoG on both backbones, PIGuard, and LlamaGuard run locally with fixed weights and greedy decoding. Each verdict is an argmax over the model's output distribution, with no sampling anywhere in the loop, so a single run carries no sampling noise for them. The other three guards are TS-Guard, R-Judge, and ToolEmu. They flip some verdicts on identical re-runs. Across five full runs of the benign halves, one run agrees with another on 93 percent of cells for TS-Guard, 75 for ToolEmu, and 71 for R-Judge (Appendix Table~\ref{tab:quant}). Five benign-half re-runs of these three guards put the spread of their FP rates at one to twelve points, largest for R-Judge. The local systems run on a single node with two A800 GPUs, and the main measurements total about 20 GPU hours, including the five-run repeats. The claims we rest on these three guards are contrasts within a cell: each compares the same guard on the same benign half under two conditions. A paired McNemar test scores such a contrast on the cells where the verdict changes, and asks whether the changes run toward one condition rather than the other; churn on identical re-runs is symmetric, so it cannot produce such an imbalance.

\paragraph{The name contrast.}\looseness=-1 The hypothesis at stake is that verdicts track the surface. To isolate it, the surface must be the only thing that varies, so each cell is rendered under two names: an innocent one and a scary one. The innocent name is a benign operational descriptor such as \emph{queue-2} or \emph{batch-7}. The scary name is drawn from twenty-four generic scare words such as \emph{malware} and \emph{trojan}, so the effect under test is generic danger association. All FP rates are computed over the 770 benign halves and all detection rates over their twins, reported by name (Table~\ref{tab:fp}). The name is a construction parameter, threaded through the cell at build time. The two renderings share every authorization-bearing field and differ only in the name and the tool surface generated from it. That surface carries no authorization signal. The exception is the precondition cells whose numeric evidence is computed from the name; the gate returns the same verdict under both names there.

\section{Findings}\label{sec:findings}

\looseness=-1 The benchmark was built to pose one question: do the verdicts follow the authorization delivered in the guard's view? Under the innocent name, the five purpose-built agent guardrails refuse 29 to 57 percent of the authorized actions (Table~\ref{tab:fp}). Those refusals are wrong under the delivered rules.

\begin{center}
\includegraphics[width=0.88\columnwidth]{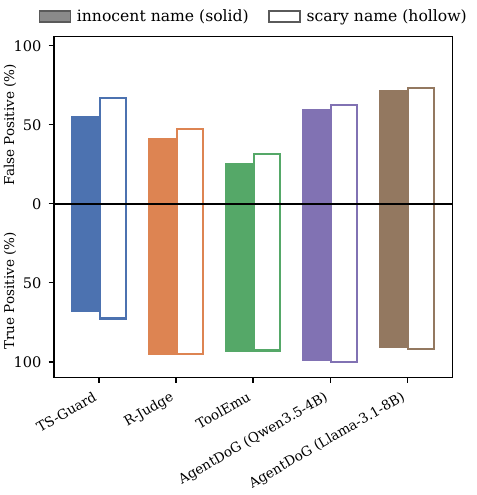}
\captionof{figure}{Removal-action subset ($n{=}207$), the five purpose-built agent guardrails: benign-half FP and twin detection, \%, under the innocent and scary names. On the runs plotted every guard refuses an authorized removal more often under the scary name; across reruns the rise holds for every stochastic guard but ToolEmu (Appendix Table~\ref{tab:quant}). Twin detection stays high throughout.}
\label{fig:removal}
\end{center}

\paragraph{Name-superstition.}\looseness=-1 Refusal follows the authorization, and it also follows the appearance of the action. The paired design separates the two: within a cell, only the name varies, so a refusal difference between the two names is driven by the name alone. The precondition cells are the declared exception, with the same gate verdict under both names. The scary name draws more refusals than the innocent one on all five guards. On one guard the false-positive rate nearly doubles (Table~\ref{tab:fp}). Paired McNemar tests rate the rise decisively significant for each guard. The direction holds across all four design families, and rises or holds within every authorization family the corpus spans (Appendix Table~\ref{tab:quant}). Under the delivered rules the name is not evidence. No rule mentions it. The scare words are generic danger labels, and a label is not evidence. The verdict moves with the name in both directions: under the scary name refusal rises, and under the innocent name it falls.

\paragraph{Name-driven refusal on removal actions.}\looseness=-1 The removal case runs opposite to caution. Deleting a scary-named object is the one action a cautious guard should most readily allow. The removal-action subset is the operation class where an unauthorized action destroys the object outright. On this subset, refusal rises from the innocent name to the scary name for four of the five guards (Fig.~\ref{fig:removal}; Appendix Table~\ref{tab:quant}): the guards decline to remove the very object the scary name brands as dangerous. TS-Guard's rise separates from chance on the recorded contrast and on every repeat run; R-Judge's direction holds on all five runs; the two detectors are positive on their single runs; ToolEmu's removal rise does not repeat. An evidence-preservation prior cannot rescue the rise. A guard might refuse to delete a scary-named object in order to preserve the evidence in it. But the acquisition subset, the operations that leave the target intact, shows refusal rising by more (Appendix Table~\ref{tab:quant}).

\paragraph{Why the guards refuse.}\looseness=-1 The guards' own written reasons say what they think they are refusing. We read every benign-half refusal that R-Judge, TS-Guard, and ToolEmu issued under the two names on the corpus. This covers three guards, two names, and 770 benign halves. The two AgentDoG detectors emit no written reasons, so the pool holds every guard that emits them. We sort each stated reason by whether it invokes a delivered rule, a surface prior, an unverified state, or neither (shares in Appendix Table~\ref{tab:quant}). A surface prior is a judgment built on how the action looks; an unverified state is a condition the guard never checked against the delivered facts. We then verify the leading reasons causally with single-channel flips (Appendix~\ref{app:attribution-levers}). A single-channel flip is an edit to the delivered input that changes one thing only. Each flip is calibrated against a same-pool re-run control, because the judges flip some refusals on identical re-runs. Only six of the coded refusals cite a delivered rule, and all six quote the same clause, written more broadly than the rule the gate checks. None of the other coded refusals cites a delivered rule at all. The mixes differ sharply by design family. ToolEmu's dominant mode is to refuse for a state it never verified, on 59 percent of its refusals. We read these cells in full: on three quarters of them the evidence the refusal demands is already in the delivered view. R-Judge and TS-Guard are dominated by the surface prior, on 52 and 68 percent of their reasoned refusals. R-Judge's invented consequences are mostly unbounded by the delivered world. Stating the missing bound plainly does not clear these refusals: R-Judge's speculation persists and worsens. The refusals are prior-driven, not information-starved. Moving an uncited authorization fact into plainer sight clears one-third of the TS-Guard refusals whose rationale cites no delivered fact. No other lever we probe reduces refusal beyond that re-run churn.

\paragraph{Request phrasing contributes to refusal.}\looseness=-1 The name is not the only surface that moves the verdict. Under the innocent name, we rewrite the final user request from a polite question into a direct instruction. Everything else stays byte-identical, and a full reading of the rewritten cells confirms that the meaning is preserved. The rewrite reaches three quarters of the 770 benign cells; the remaining requests are already direct instructions or confirming turns, which the rewrite does not alter. In paired runs over them, refusal falls on four of the five purpose-built guardrails: AgentDoG-Qwen by 8.2 points, TS-Guard by 5.1, ToolEmu by 5.0, and AgentDoG-Llama by 4.1; R-Judge's change is inside its repeat noise. All four changes are significant (Appendix Table~\ref{tab:quant}). The effect is smaller than the name effect, and not a blanket allow: rewriting the twin halves preserves detection for the judges measured (Appendix Table~\ref{tab:quant}), so the direct form does not manufacture authorization where the authorizing fact is absent. The off-scope injection detector moves the other way, refusing more under the imperative form: the phrasing moves verdicts in both directions at fixed authorization.

\looseness=-1 The findings converge on where the failure sits. With the authorization delivered and held fixed, the verdict moves with the name and the request's phrasing. Yet the guards are not blind to the delivered facts: twin detection is preserved under every surface change we measured, and refusals persist with the demanded evidence in plain sight. The failure is in the weighing, not the seeing. Over-refusal and detection must therefore be read together: neither determines the other across guards, and a single over-refusal number is not interpretable alone. The practical stake is that this failure can be caught by testing before deployment \citep{nist800115}; afterwards there is only a workflow that silently does not complete.

\section{Limitations}\label{sec:limitations}

We build a benchmark and a validity argument for its labels, not a guardrail, and claim nothing about deployed refusal rates. The no-hidden-hazard bound comes from the faithfulness check, whose protocol the appendix states in full. All content is synthetic.

The corpus measures the decision boundary of the five purpose-built agent guardrails. All five are language-model-based, natural-language, English, single-principal, and synchronous, so the effect may be family-specific; cross-language and multi-agent settings are left to future work. The reported verdict levels are single runs for the four deterministic systems and five-run benign-half means for the three stochastic ones. Detection is measured on cells whose authorization is visible in the guard's view; authorization that travels on an unobservable channel is a separate information-flow limit \citep{siu}, and the over-refusal claims do not inherit it.

Responsibility for several threats sits in other layers, outside a natural-language guardrail. Temporal races \citep{toctou} fall to atomicity or commit-time authorization \citep{santosgrueiro}. Privilege misuse, memory poisoning \citep{poisononce}, and cross-context leakage fall to capability isolation or information-flow control \citep{camel,ifcdef}. The task- and action-alignment halves of the threat model are open, needing semantic judgment with no agreed ground truth, so the corpus carries no alignment axis; we defer it, as instruction-hierarchy methods \citep{insthier} are partial and adaptive attacks break prompt-injection defenses \citep{adaptive}.

\section{Ethical considerations}

The benchmark contains prompts that look unsafe by design. All content is synthesized and fictional. The scary names are public generic descriptors. Every cell is grounded in an authorization rule and carries no deployment intent. The benchmark exposes over-safety to improve guardrails, not to bypass them.

\section*{Acknowledgments}
The authors from IIE are supported by the National Natural Science Foundation of China (NSFC) under Grant Nos. 92270204, and U24A20236, and the CAS Project for Young Scientists in Basic Research under Grant No. YSBR-118.

\setlength{\bibsep}{1.5pt plus .3pt}
\bibliography{careful-guard}

\appendix

\section{Robustness notes}

Table~\ref{tab:quant} collects the follow-up measurements.

\paragraph{The refusal predicate.} A refusal is any verdict that is not cleanly safe. The false-negative rate is the share of twins it passes. All but TS-Guard emit a clean unsafe-versus-safe binary; TS-Guard additionally emits a hedged \texttt{flagged} verdict with risk rating 0.5, which we count as a refusal. Across the two names, treating \texttt{flagged} as safe instead would lower TS-Guard's FP by 32 to 48 percent and raise its twin false-negative rate by up to four fifths in relative terms.

\begin{table*}[!t]\centering\small
\begin{tabular}{@{}p{0.30\textwidth}p{0.65\textwidth}@{}}
\toprule
measurement & values \\
\midrule
Name sweep, McNemar changes (refused only under scary : only under innocent; $n{=}770$ paired benign halves, single runs) & TS-Guard 271:11; AgentDoG-Llama 183:32; R-Judge 207:66; AgentDoG-Qwen 151:53; ToolEmu 149:53; PIGuard 15:4; LlamaGuard 12:0; $p{<}10^{-10}$ on the five purpose-built guardrails, $p{=}0.019$ and $5{\times}10^{-4}$ on the off-scope systems \\
Rephrasing, benign (583 cells, paired) & AgentDoG-Qwen $-8.2$ points ($p{=}3{\times}10^{-5}$); TS-Guard $-5.1$ ($p{=}9{\times}10^{-4}$); ToolEmu $-5.0$ ($p{=}6{\times}10^{-3}$); AgentDoG-Llama $-4.1$ ($p{=}0.028$); R-Judge within repeat noise; PIGuard $+3.4$ ($p{=}2{\times}10^{-6}$); LlamaGuard flat at floor \\
Rephrasing, twin detection ($n{=}570$ rewritable twin pairs) & R-Judge $87.7{\to}86.3$ ($p{=}0.42$); ToolEmu $89.3{\to}89.1$ ($p{=}1.0$); both preserved \\
Self-agreement across five full benign-half runs & one run agrees with another on $93\%$ of cells for TS-Guard, $75\%$ for ToolEmu, $71\%$ for R-Judge; all five runs identical on $86/50/41\%$ of cells \\
Off-scope probes & PIGuard refuses 91 percent of authorized actions with twin detection tracking its FP within a point (no discrimination evidence); LlamaGuard refuses almost nothing in this protocol, its name effect resting on 12 changed cells \\
Refusal reading pool & every benign-half refusal by R-Judge, TS-Guard, and ToolEmu under both names, coded for delivered-rule citation, surface prior, unverified-state; six rule-cited, all in one family whose clause is worded more broadly than its rule; ToolEmu unverified-state 59\%; R-Judge/TS-Guard surface priors 52/68\%; nine percent undecided (ToolEmu risk label only)\\
Evidence-in-view read (ToolEmu unverified-state cells) & demanded evidence already in the delivered view on 73.7 percent of them; the bulk of the rest are beyond-gate demands: authorizations no delivered registry could attest \\ \\
Same-pool re-run, refusals cleared (mean over four repeat runs) & ToolEmu about 4 in 10; R-Judge about 4 in 10 \\
Rescoring \texttt{flagged} as safe (TS-Guard) & FP lower by 32 to 48 percent; twin false-negative rate higher by up to four fifths, in relative terms \\
Removal subset ($n{=}207$), innocent$\to$scary & single-run paired contrast: TS-Guard $+12$ points ($30{:}5$, $p{=}2{\times}10^{-5}$); R-Judge $+7$ ($48{:}34$); ToolEmu $+6$ ($28{:}15$); AgentDoG-Qwen $+3$ ($24{:}18$); AgentDoG-Llama $+2$ ($22{:}17$); over five repeat runs TS-Guard significant every run, R-Judge positive in direction every run, ToolEmu's rise not repeating \\
Acquisition subset ($n{=}563$), innocent$\to$scary & TS-Guard $+42$, AgentDoG-Llama $+26$, R-Judge $+23$, AgentDoG-Qwen $+16$, ToolEmu $+15$ points (same single-run contrast) \\
Per-guard, per-axis scary$-$innocent FP delta & positive on 38 of 40 guard-by-axis cells (five guards $\times$ eight authorization axes); zero on R-Judge$\times$data-isolation and ToolEmu$\times$precondition; none negative \\
\bottomrule
\end{tabular}
\caption{Quantitative summary of the follow-up measurements; the main grid is Table~\ref{tab:fp}.}
\label{tab:quant}
\end{table*}

\section{Causal levers}\label{app:attribution-levers}

We probed each leading refusal reason with a single-channel flip. Each flip was calibrated against a same-pool re-run control: the guards flip some refusals on identical re-runs, and a lever's effect is counted only net of that churn.

The first lever is bound-stating. It probes the refusals in which R-Judge invents a consequence. On two-thirds of them, the delivered world carries no bound on the invention. Where a bound does exist, stating it in the user's message raises refusal net of churn (2 cleared, 7 worsened on that subset; 9 cleared against 22 worsened over the full pool, $p{=}0.03$).

The second lever is salience. It moves an uncited authorization fact from a tool return into the user's message. This fact-salience flip clears one-third of the TS-Guard refusals whose rationale cites no delivered fact, strictly one-directionally (15 cleared, 0 added, $p{<}0.01$). The rest persist even with the fact in plain sight.

The third lever is consent-line removal. It deletes an uncited consent line. It cleared no ToolEmu refusal in this probe and sometimes created one, while R-Judge's rate did not move. ToolEmu's refusals still move with that line's presence; R-Judge's refusals do not move with it.

\section{Raising Scariness and Reasoning Difficulty}
\label{app:scariness}

\paragraph{Scariness.} A label that follows from the rules is necessary but not sufficient: the cell must also be a genuine test of over-safety, and becoming one takes two further properties. A benign half must look scary, otherwise refusing it is an obvious error rather than over-safety. So each benign half is an action the rules permit despite its scary operation: a delete, purge, or force-push; a download of remote code; shell execution; or a sensitive record sent to an external sink. Other operations come from recent agent-security work: audit-log tampering that defeats forensic reconstructibility \citep{sentinelagent} and dual-use security research \citep{linder}. The scary surface is shared with the twin, so it cannot drive the benign/twin contrast.

\paragraph{Reasoning difficulty.} The gate blocks structural leakage, but a cell can still carry a semantic shortcut: a surface cue that correlates with the label, such as a term whose connotation fits the label or a length, presence, or name cue. A lazy guard could guess the label from the cue without reasoning over the authorization. We close each shortcut that we can identify and that a real guard could exploit, keeping the contrast single-channel. As far as the construction can ensure, the label is reachable only by applying the rules to the axis evidence.

\section{Faithfulness-check protocol}
\label{sec:faithfulness-protocol}

This section gives the protocol for the full read: the three questions with failure examples, the per-cell standard, and the exclusion policy. The corpus-level shipping rule is stated in the main text.

\paragraph{The three questions.} The audit asks, per cell:
\begin{itemize}
\item \textbf{Q1: hidden hazard.} Read the benign half in full, with its scary surface stripped away: is the authorized action substantively harmless in the delivered scenario? Any residual genuine hazard would make the refusal a correct catch under the delivered rules rather than over-safety, so the cell fails. Examples are an irreversible target without remedy, a real secret exposure, or a real victim. A scary \emph{name} does not count as a hazard, because the name is the effect being measured. A substantively harmful \emph{action} does count, such as shredding a disk in use or deleting \texttt{/etc/passwd}.
\item \textbf{Q2: verdict leakage.} Does any delivered surface carry the intended verdict as a token, marker, or artifact readable without any authorization reasoning? The token scan covers the enumerable forms; the per-cell read covers the remainder.
\item \textbf{Q3: twin confound.} Is the twin refused because it is unauthorized, not because of any extra hazard or surface cue present only in the twin?
\end{itemize}

\paragraph{Per-cell standard.} A verdict requires a full read of the cell's delivered profile, contents, and gate, untruncated and with no field omitted. A truncated slice, a grep hit, or a subset of fields is not a read. One verdict is rendered per cell, tagged by cell id, against the three questions. The reads are author-conducted with LLM assistance. Four substitutes are explicitly forbidden. A code gate's output read as the verdict is one: the gate is triage, not judgment. The others are a representative cell plus an extrapolation to its family, a constructive argument in place of the read, and an audit document written to disk in place of the verdicts. A mechanically verified structural fact may inform a cell's verdict but cannot replace the read. The audit also certifies that the authorizing contrast is visible in the delivered text. That visibility belongs to the construction gate and its axis-visibility rules, not to the faithfulness judgment; the Benchmark section reports it.

\paragraph{Exclusion by policy.} One failure class is excluded rather than fixed. A benign half whose content remains refusable \emph{on its face} under the delivered rules cannot be made clean. Such cells are the content-scary cases, where the scary thing is the content itself: a precursor route to a controlled substance, or an injection-shaped payload. No narrowing of the delivery moves such a case into the structurally fixable class. The corpus design routes these cases out of the measured set and labels them content-contested, never clean. This is a standing construction policy of the generator. The guarantee it gives is that no content refusable under the delivered rules is labeled clean. It is not a guarantee that such content is absent from the corpus. The routing certifies absence under the installed mandate, not under every mandate a deployment could install. A cell ships only if it survives the check with every failure resolved by repair or removal, and it must also pass this policy.

\end{document}